# A Commitment-Based Hybrid Post-Quantum Cryptographic Model for Multi-File Cloud Storage

**Lemdi Frank Prikutse, Regina Esi Turkson, Alimatu-Saadia Yussiff, Abdul-Lateef Yussiff, Maame G. Asante-Mensah**

## ABSTRACT

Cloud storage clients increasingly require authentication that remains secure against future quantum-capable adversaries, motivating hybrid constructions that combine classical primitives with standardized post-quantum alternatives. Extended naively to multi-file upload, such constructions incur a per-file lattice signing cost that dominates authentication time and becomes prohibitive at realistic batch sizes. This paper presents a commitment-based hybrid post-quantum model that addresses this bottleneck. It comprises AES-256-GCM bulk encryption, a hybrid X25519 with ML-KEM-768 key encapsulation mechanism, and a hybrid Ed25519 with ML-DSA-65 dual signature, computed over a SHA3-256 batch commitment. The commitment binds all ciphertexts in a batch to a single fixed-size digest that is signed once, reducing the number of post-quantum signature invocations per batch from $n$ to one, independent of batch size; the remaining encryption and hashing is bounded by fast symmetric throughput. On a commodity client platform, averaged over 20 repetitions, this holds signing-phase time near-constant as the batch grows while the per-file baseline scales linearly. At $n = 1000$, the model reduces signing-phase time by factors of 629, 606, and 725 for 100 KB, 1 MB, and 10 MB files respectively, against a per-file dual-signing baseline sharing every other primitive.

## 1. INTRODUCTION

Cloud computing has emerged as the dominant paradigm for data storage and processing, with organizations migrating critical workloads to cloud infrastructure at unprecedented scale [1]. However, this centralization creates significant security challenges concerning long-term confidentiality and authenticity [2-3]. Current cloud cryptographic systems rely on classical primitives including Rivest Shamir Adleman (RSA), Elliptic Curve Cryptography (ECC), and Advanced Encryption Standard (AES), which provide adequate security against conventional threats but face serious threats emerging from advances in quantum computing. The polynomial-time efficiency of Shor's algorithm in addressing integer factorization and discrete logarithm challenges on quantum hardware threatens the cryptographic security of extensively used systems including RSA, Diffie-Hellman, and ECC [4-6].

The National Institute of Standards and Technology (NIST) Post-Quantum Cryptography (PQC) standardization process finalized the first quantum-resistant algorithms, including Module-Lattice-Based Key Encapsulation Mechanism (ML-KEM) and Module-Lattice-Based Digital Signature Algorithm (ML-DSA) [7-8].

Data encrypted today using classical algorithms remain vulnerable to Harvest Now Decrypt Later (HNDL) attacks, where adversaries gather encrypted data to decrypt once quantum computers become available, particularly threatening cloud-stored data requiring long-term confidentiality, including government communications, medical records, and financial information [9].

However, transitioning entirely to Post-Quantum Cryptography (PQC) algorithms faces several challenges: limited real-world deployment experience compared to battle-tested classical cryptography [10], significant investment and operational complexity in migrating existing cloud infrastructure including software re-engineering and infrastructure upgrades [11], and uncertainty about the long-term security of specific PQC constructions as cryptanalytic advances continue [10]. Hybrid cryptographic approaches which combine classical and post-quantum primitives offer a pragmatic path forward, maintaining security if at least one component remains unbroken and providing defense-in-depth against both known and unknown cryptanalytic advances [12].

The practical integration of these hybrid primitives into multi-file cloud storage workloads, however, exposes a performance barrier that has received less attention than the cryptographic design problem itself. Lattice-based signatures are large. ML-DSA-65 produces signatures of approximately 3.3 KB [13], more than fifty times the size of an Ed25519 signature [14]. In a naïve per-file authentication scheme, the number of ML-DSA-65 signing operations grows linearly with $n$, so that the total signing-phase cost is dominated by the lattice signing operations rather than by the bulk AES-256-GCM encryption.

This paper therefore addresses the multi-file signing-cost problem through a commitment-based design. The paper makes the following contributions:

A post-quantum hybrid cryptographic model tailored for multi-file cloud upload authentication. The construction uniquely combines AES-256-GCM for bulk encryption, X25519/ML-KEM-768 for hybrid key encapsulation, and Ed25519/ML-DSA-65 dual signatures bound to a pre-signature SHA3-256 batch commitment.

Secondly, this paper demonstrates that aggregating files into a single batch commitment decouples signature generation from batch size, reducing lattice-based signature (ML-DSA-65) invocations from $n$ to 1 per batch while preserving the provable security properties of the constituent primitives.

Finally, this paper evaluates implementation on constrained commodity hardware (Intel Core i5-6300U). For $n = 1{,}000$, the proposed model yields signing-phase speedups of 629 times, 606 times, and 725 times across file sizes of 100 KB, 1 MB, and 10 MB respectively, compared to a per-file dual-signing baseline.

## 2. RELATED WORK

The AES-ECC hybrid has emerged as a widely adopted architecture for cloud data protection due to ECC's compact key representation and computational efficiency[15]. However, ECC's security rests on the Elliptic Curve Discrete Logarithm Problem (ECDLP) [16] which Shor's quantum algorithm takes polynomial time to solve [17], while AES remains vulnerable to Grover's algorithm that reduces its effective security level [18]. The Harvest Now Decrypt Later (HNDL) threat renders AES-ECC inadequate for long-term confidentiality, creating urgent demand for post-quantum alternatives that maintain comparable performance [9].

However, hybrid constructions combining post-quantum and classical primitives offer defense-in-depth during quantum transition [19]. Concatenation-based combiners have emerged as the provably secure approach for hybrid Key Encapsulation Mechanisms (KEMs), where shared secrets and ciphertexts from multiple KEMs are concatenated before key derivation, retaining Indistinguishability under Chosen-Ciphertext Attack (IND-CCA) security, provided a minimum of one component remains secure [20]. This robustness enables cryptographic hedging against both inadequately analyzed post-quantum algorithms and quantum-vulnerable classical schemes, while simpler XOR-based alternatives fail to preserve Chosen-Ciphertext Attack (CCA) security [20]. Deployment studies of hybrid post-quantum Transport Layer Security (TLS) report 20%-40% handshake latency increase primarily from larger key sizes rather than computational overhead [21], though comprehensive performance evaluation of hybrid protocols for bulk data encryption in cloud storage remains limited.

NIST's standardization has produced ML-KEM responsible for key encapsulation and ML-DSA responsible for digital signatures, with contemporary benchmarking demonstrating competitive performance compared to classical systems[22]. Performance comparisons show that ML-KEM encapsulation and decapsulation operations are faster than RSA and comparable to ECDH, while ML-DSA signing is comparable to Elliptic Curve Digital Signature Algorithm (ECDSA).

This paper's proposed model adopts an AND-combiner for its signature layer (Ed25519 $\wedge$ ML-DSA-65) and a concatenation-based hybrid KEM (X25519 $\wedge$ ML-KEM-768). The novelty lies not only in the hybrid construction itself, which follows the framework from [23], [24], but also in its interaction with the multi-file batch authentication mechanism.

While prior work has explored hybrid signature schemes for Transport Layer Security (TLS) and authentication protocols [21], [25] theoretical foundations of hybrid post-quantum cryptography [26], hash-based commitment schemes for digital signatures [27], [28], and performance overhead of post-quantum cryptography in network protocols [21], these approaches primarily target protocol handshakes, small messages, or theoretical frameworks rather than the specific challenge of signing large encrypted files in cloud storage systems where file sizes reach hundreds of megabytes [29].

Furthermore, while commitment schemes are established in cryptographic protocols [27], [28], and hybrid classical-PQC systems have been proposed for various applications [21], [24], [25], [26], [30] their combination for performance optimization in multi file cloud storage cryptography has not been extensively studied. The workload-level question of how hybrid signatures should be efficiently applied to a batch of files at the scales encountered in cloud object storage needs to be addressed.

A lattice-based signature aggregation scheme was proposed by [31], achieving constant-time verification after $O(n)$ aggregation for low-latency distributed networks such as sensor systems, smart grids, and blockchain platforms. Although their objective was to reduce per-message cost of post-quantum authentication, their target was fundamentally different. Their scheme aggregates individual signatures produced by many independent signers, and the constant-time benefit accrues to verification after per-signer signatures have already been computed, whereas cloud multi-file uploads involve a single signer for whom the dominant cost is the signing operation itself.

In another work, [32] present a blockchain-integrated framework combining lattice-based post-quantum primitives with distributed ledger technology to protect the privacy and integrity of multimedia data in cloud-enabled public auditing platforms. Their work explicitly motivates the adoption of post-quantum primitives by the harvest-now-decrypt-later threat to long-lived cloud data and demonstrates the feasibility of integrating lattice-based cryptography into cloud auditing workflows. Nevertheless, their optimization target is the integrity assurance of the auditing platform under a blockchain trust model rather than the client-side cost of authentication at upload time, and their construction does not treat the hybrid classical-plus-post-quantum composition through which the present work hedges transition-period risk.

Two lines of work approach the batch-signing problem more directly. [33] propose lattice-based batch signature schemes in which a single signer authenticates a group of messages at approximately the cost of one signature operation, with signing and verification complexity independent of the number of messages $k$. Their constructions employ either a binary-tree structure or an intersection method over lattices, and are proven existentially unforgeable under adaptive chosen-message attack based on the hardness of the Small Integer Solution problem. Their target, however, is message integration authentication in IoT sensor networks, their analysis is asymptotic rather than empirical, and their schemes are purely lattice-based, offering no classical component against which to hedge post-quantum cryptanalytic risk during the transition period.

[34] come closer to the present setting, combining classical and post-quantum signatures under a nested combiner of the form employed in [23], and detaching commitment computation from per-item signing to reduce the cost of authenticating large batches. Their efficiency, however, depends on a hardware-assisted commitment construction executed within a trusted execution environment, and their deployment context is continuous data offload for digital twins and low-end IoT devices rather than client-side multi-file upload to cloud object storage.

In essence, the batch-signing mechanism has been established for purely post-quantum schemes [33] and, in hybrid form, for hardware-assisted streaming workloads [34], while hybrid classical-post-quantum composition [23], [24] and post-quantum deployment in cloud storage [32] have advanced separately. To the best of our knowledge, no prior work combines a hybrid classical-post-quantum signature with a pre-signing batch commitment, realized in software without secure-hardware assumptions and evaluated empirically across batch sizes, for single-client multi-file upload to cloud storage. This is the gap the present work addresses.

## 3. PROPOSED MODEL DESIGN

### 3.1 Cryptographic Preliminaries

*Hybrid Key Encapsulation*

The hybrid KEM combines Curve25519 and ML-KEM-768 through concatenation followed by key derivation. The encapsulation process generates ephemeral Curve25519 keys, performs ML-KEM-768 encapsulation, concatenates both shared secrets ($K_{X25519} \parallel K_{ML-KEM-768}$), and applies HKDF-SHA-256 to derive the final symmetric key $K_s$. This concatenation-based approach, when combined with HKDF, ensures that the derived key maintains full security as long as a minimum of one KEM (either Curve25519 ECDH or ML-KEM-768) remains secure against attacks. This provides defense-in-depth against cryptanalytic advances.

*Digital Signatures*

Authentication uses AND-composed dual signatures. Both Ed25519 and ML-DSA-65 signatures are generated over the same message, and verification succeeds only if both signatures are valid. This composition ensures that forging a signature requires breaking both schemes, which provides authentication security that gracefully transitions to post-quantum security.

Our proposed model integrates five carefully selected cryptographic algorithms:

*AES-256-GCM:* Authenticated encryption provides confidentiality and integrity for bulk data. The 256-bit key size maintains security even with Grover's quantum search algorithm, which reduces effective security to 128 bits. Combined with ML-KEM-768's quantum resistance, the overall system achieves 128 bits of post-quantum security.

*Curve25519 ECDH:* Ephemeral Elliptic Curve Diffie-Hellman provides forward secrecy and efficient classical key exchange. It offers approximately 128 bits of classical security.

*ML-KEM-768:* NIST-standardized lattice-based key encapsulation mechanism provides post-quantum key exchange with NIST security level 3, 128 bits quantum security.

*Ed25519:* Elliptic curve digital signature provides efficient classical authentication with 128 bits of classical security.

*ML-DSA-65:* NIST-standardized lattice-based digital signature provides post-quantum authentication with NIST security level 3.

### 3.2 Architecture of Proposed Hybrid Model

The proposed model adopts a symmetric sender-receiver architecture, in which both parties share an identical stack of modules: a Key Management and Crypto Engine, a Data Encryption/Decryption Module, a Commitment Module, and an Authentication and Validation Module. The sender passes plaintext files through these modules to produce a sealed batch of ciphertexts, a single SHA3-256 commitment, and a dual signature (Ed25519 and ML-DSA-65), which is transmitted over a public channel. The receiver reverses the process, accepting the batch only if the recomputed commitment matches and both signatures verify.

Fig. 1 below shows a detailed architecture of the proposed AES-ECC-PQC hybrid cryptographic model.

*Key Management and Crypto Engine:* The architecture incorporates a Key Management and Crypto Engine responsible for key generation, exchange, encapsulation, and derivation at both communicating parties (A and B). The Elliptic Curve Cryptography (ECC) Engine implements X25519 (Curve 25519) for Elliptic Curve Diffie-Hellman (ECDH) key exchange, selected for strong security guarantees, compact key sizes of 32 bytes, and low computational overhead suitable for high-performance scalable cloud systems. In parallel, the Post-Quantum Key Encapsulation Mechanism (PQ-KEM) based on ML-KEM-768 performs key encapsulation and decapsulation to generate a post-quantum shared secret, ensuring resilience against quantum attacks including Shor's algorithm which threatens classical public-key cryptography. Classical and post-quantum shared secrets are concatenated and processed through Hash-based Message Authentication Code (HMAC)-based Key Derivation Function (HKDF) using SHA-256, which performs extract-and-expand operations to produce a high entropy 256-bit symmetric session key with strong unpredictability. This hybrid key derivation approach ensures security as long as a minimum of one underlying cryptographic assumption holds. This achieves both forward secrecy and quantum resilience through defense-in-depth. Per-session nonce and salt generation strengthen key entropy, ensure freshness and uniqueness of cryptographic operations, prevent replay attacks, and protect against precomputation and related-key vulnerabilities.

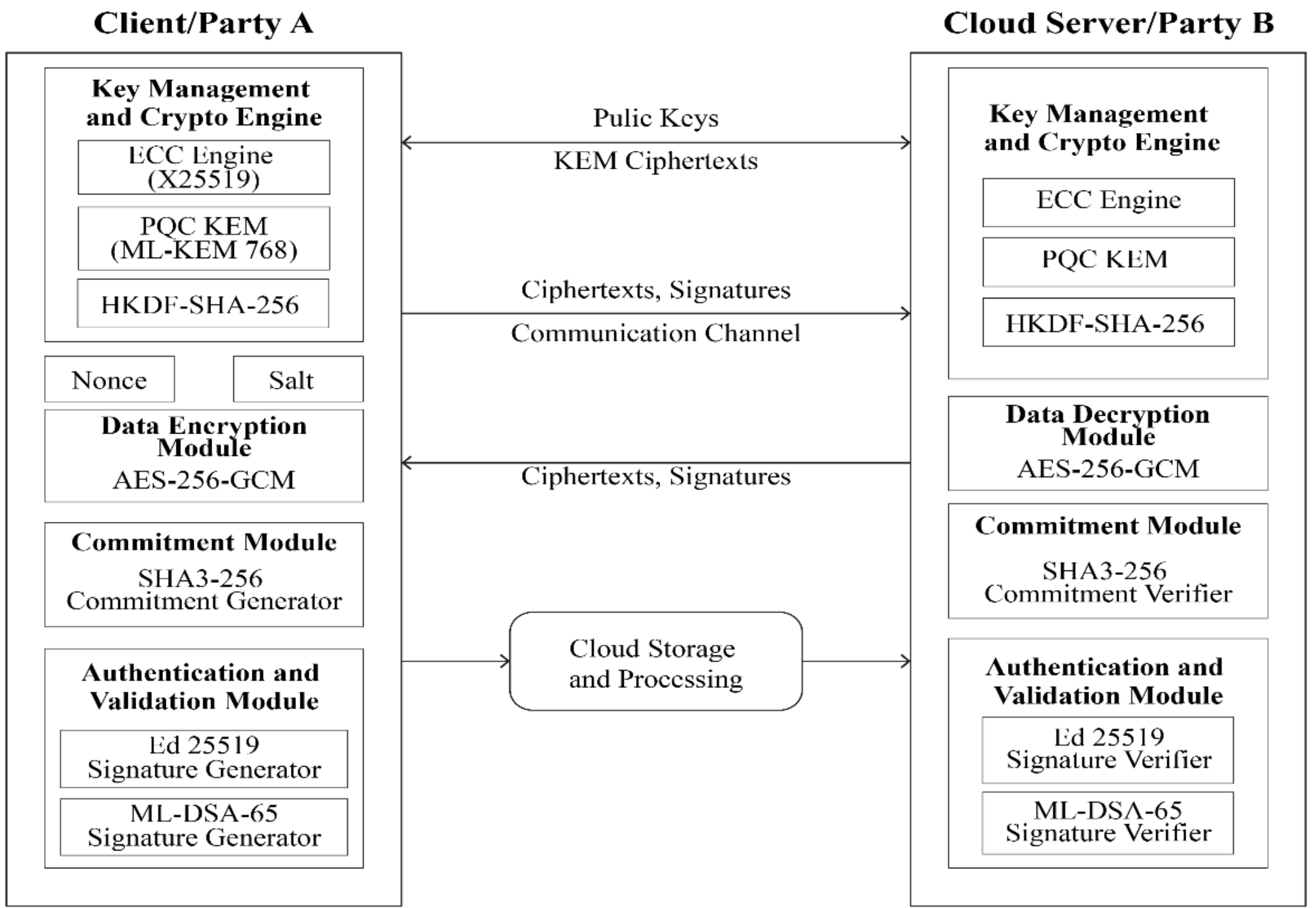

***Fig.1:*** *Detailed Architecture of Proposed AES-ECC-PQC Model*

*Data Encryption Module:* This module ensures data confidentiality and integrity using the derived symmetric session key. Encryption employs Advanced Encryption Standard in Galois Counter Mode (AES-256-GCM), chosen for strong security properties, authenticated encryption capability, and implementation efficiency. GCM provides confidentiality, integrity, and authenticity in a single operation, reducing computational overhead. As shown in Fig. 1, encrypted data flows between the Data Encryption Module and Cloud Storage, remaining unreadable and tamper-evident even if infrastructure or communication channels are compromised.

*Data Decryption Module:* The Data Decryption Module reconstructs the plaintext from received ciphertexts using AES-256-GCM under the session key derived by the receiver's Key Management and Crypto Engine. For each ciphertext in the batch, the module recomputes the GCM authentication tag and decrypts only if the tag verifies, ensuring per-file integrity and authenticity at the symmetric layer. Any tampering with a ciphertext, its associated data, or its nonce causes tag verification to fail and the affected file is rejected without decryption. Successful decryption yields the original plaintext which is forwarded only after the Commitment Verification and Authentication modules also confirm batch-level integrity.

*Commitment Module:* The Commitment Module produces a single cryptographic binding over the entire batch of file ciphertexts using SHA3-256. This commitment captures the integrity of every ciphertext in the batch in one short, fixed-size value, which is then included in the transcript signed by the Authentication and Validation Module. Any tampering (substitution, reordering, or modification) with any ciphertext causes the recomputed commitment at the receiver to mismatch and the batch to be rejected. By replacing per-file signing with a single signature over the batch commitment, the module reduces the number of post-quantum signature operations from $O(n)$ to $O(1)$ thus from $2n$ invocations to two, irrespective of batch size, while preserving integrity guarantees.

*Authentication and Validation Module:* This module employs hybrid digital signatures for entity authentication, integrity verification, and non-repudiation. Party A generates dual signatures using Ed25519 (classical elliptic curve authentication) and ML-DSA-65 (lattice-based post-quantum signature) over cryptographic transcripts. Party B verifies both signatures through corresponding Ed25519 and ML-DSA-65 verifiers. Successful dual verification confirms message legitimacy and session integrity, ensuring authentication security even if either classical or post-quantum schemes are compromised.

*Cloud Storage and Processing:* This component represents the external cloud environment positioned outside the cryptographic trust boundary. All data entering this component is protected by AES-256-GCM, with authentication and validation performed before acceptance. This design ensures plaintext data and cryptographic keys are never exposed to cloud infrastructure.

The proposed hybrid post-quantum cryptographic model operates through a number of steps which ensure both classical and quantum security:

(1) The sender performs X25519 Elliptic Curve Diffie-Hellman (ECDH) key exchange and ML-KEM-768 key encapsulation with the recipient's public keys to generate dual shared secrets, providing defense-in-depth against both classical and quantum adversaries.

(2) Both secrets are concatenated and processed through HKDF-SHA-256 to derive a high-entropy 256-bit AES session key, ensuring security as long as at least one underlying primitive remains unbroken.

(3) The plaintext is encrypted using AES-256-GCM with Additional Authenticated Data (AAD) to produce ciphertext, providing confidentiality, integrity, and authenticity in a single cryptographic operation.

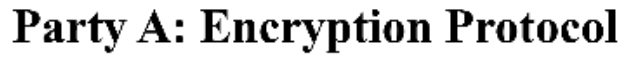

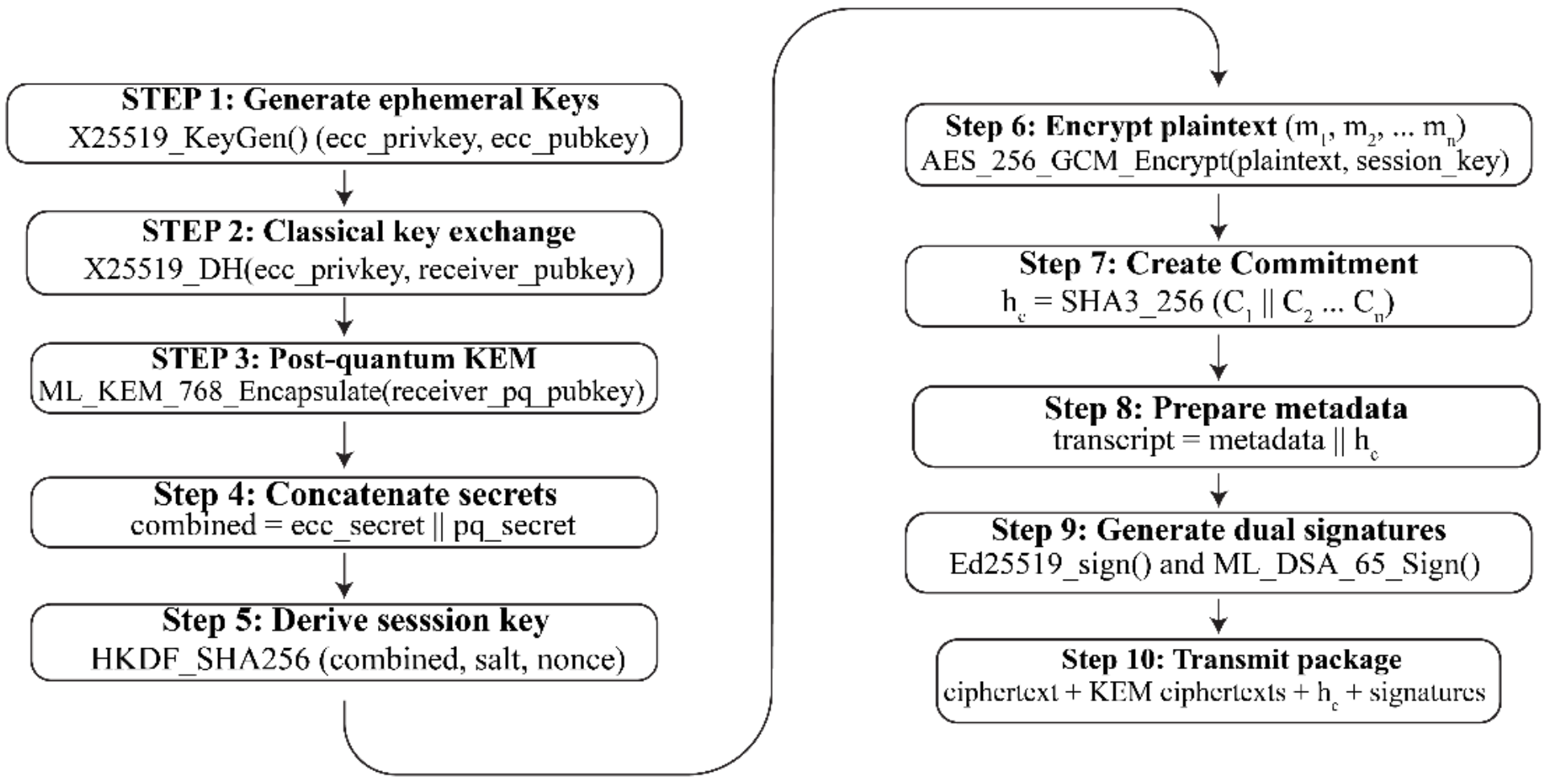

***Fig.2:*** *Encryption Protocol Flow*

**Party B: Decryption Protocol**

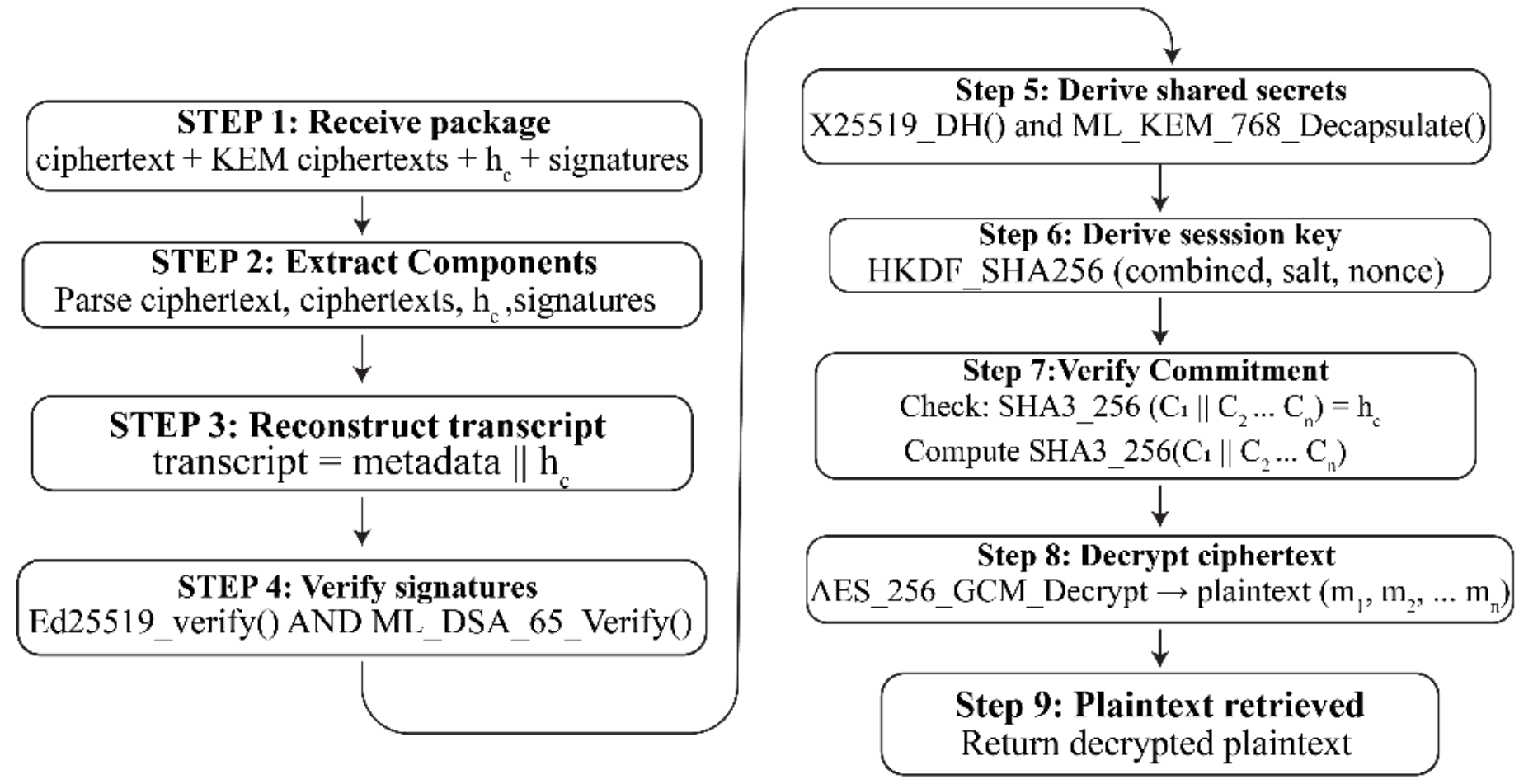

***Fig. 2:*** *Decryption Protocol Flow*

(4) A cryptographic commitment $h_c$ is computed over the full set of batch ciphertexts using the SHA3-256 hash function, binding all $n$ ciphertexts into a single fixed-size value that is subsequently signed once, rather than signing each ciphertext individually.

(5) A cryptographic transcript T is constructed containing the nonce, salt, authentication tag, ML-KEM ciphertext, AAD, and commitment.

(6) Dual signatures using Ed25519 and ML-DSA-65 are generated over T, providing hybrid authentication resistant to both classical and quantum attacks.

(7) The complete package containing ciphertext, cryptographic parameters, and both signatures is transmitted to the recipient.

(8) The recipient reconstructs transcript T and verifies both signatures for authenticity.

(9) Performs ECDH and ML-KEM decapsulation operations to recover the identical session key via HKDF.

(10) Decrypts the ciphertext using AES-256-GCM with AAD verification.

(11) Validates that the recomputed commitment matches the received commitment before outputting the plaintext, while rejecting any package with invalid signatures or mismatched commitments to prevent forgery and tampering attacks.

Fig.2 and Fig.3 above show the encryption and decryption protocols respectively.

### 3.3 Formulation of the Proposed Model

The proposed hybrid post-quantum cryptographic model ($\Pi$) is represented as follows:

$$\Pi = (Sign_{(ECC,PQC)}) \circ (AEAD_{HKDF(ECDH \| KEM)}) \quad (1)$$

$$\Pi = (\sigma_{Ed25519} \wedge \sigma_{ML-DSA-65}) \circ (AES - GCM_{K_S}) \quad (2)$$

$$\Pi = (\sigma_{Ed25519}(sk_A, T) \wedge \sigma_{ML-DSA-65}(sk_A, T)) \circ (AES - GCM_{K_S}) \quad (3)$$

To optimize signature performance, a cryptographic commitment, $h_c$ to the ciphertext $C$ is created:

$$h_c = \text{SHA3-256}\ ((C_1) \| \cdots \| (C_n)) \quad (4)$$

Where,

$K_s$ = HKDF-SHA-256 ($K_{X25519} \| K_{ML-KEM-768}), S, N$ )

$$K_{X25519} = ECDH(\, sk_A \,, pk_B)$$

$$K_{ML-KEM-768} = Encaps(\, pk_B)$$

$$T = N \| S \| \tau_n \| C_{PQC} \| AAD \| h_c$$

$\sigma_{Ed25519}$ = classical signature

$\sigma_{ML-DSA-65}$ = post quantum signature

$C$ = $batch$ ciphertext (AES − GCM encrypted data)

$C_{PQC}$ = ciphertext (AES − GCM encrypted data)

$sk_A$ = the private key of Party A

$pk_B$ = the public key of party B.

$T$ = the transcript ( the data being signed)

$AAD$ = Additional Authenticated Data

$h_c$ = ciphertext commitment

$S$ = salt for HKDF

$N$ = nonce for AES − GCM

$\tau_n$ = authentication tag for AES − GCM

The proposed model is justified by showing that each security goal is supported by a primitive that resists both classical and quantum adversaries, with classical components paired against post-quantum counterparts so that no single break compromises the system.

*Confidentiality:* Payload data is protected by AES-256-GCM, which retains approximately 128 bits of security under Grover's algorithm. The session key is derived from a hybrid combination of X25519 and ML-KEM-768 through HKDF-SHA-256, so recovery requires breaking both the elliptic curve discrete logarithm problem and the Module-LWE problem simultaneously.

*Integrity:* AES-GCM's authentication tag detects bit-level tampering at the level of the individual ciphertext. At the batch level, the SHA3-256 commitment $h_c$ = SHA3-256 $((C_1) \| \cdots \| (C_n))$ binds the entire ordered batch into a single 32-byte verifiable value: because concatenation is order-sensitive and the transcript records the file count $n$, the commitment protects against individual file tampering, file reordering, insertion of unauthorised files, and deletion of legitimate files. The commitment reduces the number of post-quantum signature operations per batch from $O(n)$ to $O(1)$, without weakening the per-file integrity guarantees provided by AES-GCM.

*Authentication:* The commitment $h_c$ is signed in parallel with Ed25519 and ML-DSA-65, and the composite signature is accepted only if both individual signatures verify. Forgery therefore requires breaking both the classical and the post-quantum signature scheme; under a quantum adversary that breaks Ed25519, unforgeability is preserved because ML-DSA-65 remains post-quantum secure. This AND-combiner preserves EUF-CMA security whenever at least one component signature scheme remains unforgeable.

*Forward secrecy and HNDL resistance*: Session keys are derived from ephemeral X25519 and ML-KEM-768 keypairs that are erased on session close; long-term keys are used only for signing. An adversary that archives today's ciphertexts for future decryption against a quantum-capable opponent gains no advantage, because session-key recovery still requires breaking ML-KEM-768, which is designed to resist quantum attack.

*Cloud service provider resistance:* The cloud service provider sees only the ciphertext, the GCM tag, the commitment, and the dual signature, none of which leaks plaintext or key material. Any tampering invalidates the GCM tag, the commitment, or the dual signature, all of which are bound together in the signed transcript.

In summary, every security goal is supported by at least one scheme that remains secure against both classical and quantum adversaries, and the commitment-based design holds the number of post-quantum signature operations constant at two per batch, irrespective of $n$, without weakening any security property.

## 4. COMPLEXITY ANALYSIS

This section derives the aggregate cost of the proposed model $\Pi$ and compares it against a per-file signing baseline that uses the same primitives but signs each file individually. The number of post-quantum signature operations required to authenticate a batch of $n$ files is reduced from $O(n)$ in the baseline to $O(1)$ in $\Pi$. The total signing-phase work remains $O(n)$ because of the commitment construction, but the per-file marginal coefficient is dominated by SHA3-256 hashing over 32-byte digests rather than by lattice signing.

### 4.1 Notation

Let $n$ denote the number of files in the batch, $|m_i|$ the size of the $i$-th plaintext, $N = \sum_{i=1}^{n}|m_i|$ the total plaintext size, and $|C_i| = |m_i| + 16$ the ciphertext size including the AES-GCM tag. All cryptographic primitives are assumed to run in time proportional to their input length, as specified by FIPS 197 for AES-256-GCM [35], FIPS 202 for SHA3-256 [36], FIPS 203 for ML-KEM-768 [37], FIPS 204 for ML-DSA-65[38], RFC 7748 for X25519 [39], RFC 8032 for Ed25519 [14], and RFC 5869 for HKDF-SHA-256 [40].

### 4.2 Per Operation Cost

The proposed scheme performs the following operations per batch, grouped by phase.

***Table 1:*** *Per Operation Complexity of Proposed Model*

| No. | Operation | Phase | Complexity |
|---|---|---|---|
| 1–3 | X25519 ECDH, ML-KEM-768 encapsulation, HKDF | Key establishment | $O(1)$ |
| 4 | AES-256-GCM encryption of the batch | Bulk encryption | $O(N)$ |
| 5 | Per-file SHA3-256 hashing of ciphertexts | Commitment | $O(N)$ |
| 6 | Outer SHA3-256 over concatenated 32-byte digests | Commitment | $O(n)$ |
| 7 | Ed25519 and ML-DSA-65 signing of the transcript | Authentication | $O(1)$ |

Operations 1-3 and 7 have fixed-length inputs and are $O(1)$. Bulk encryption and per-file hashing scale with the total ciphertext size $N$. The outer commitment hashes $32n$ bytes and scales with $n$ only.

### 4.3 Aggregate Complexity

Summing per-operation costs, the total cost of $\Pi$ for a batch of $n$ files totalling $N$ plaintext bytes is:

$$T_{\Pi}(n, N) = c_{\text{KEM}} + (c_{\text{AES}} + c_{\text{H}}) \cdot N + 32c_{\text{H}} \cdot n + c_{\text{Sig}} \tag{5}$$

where $c_{\text{KEM}}$ is the fixed key-establishment cost, $c_{\text{AES}}$ and $c_{\text{H}}$ are per-byte AES-256-GCM and SHA3-256 rates, and $c_{\text{Sig}}$ is the fixed cost of the Ed25519 + ML-DSA-65 dual signature. The corresponding baseline cost, in which each file is signed with its own dual signature, is:

$$\boldsymbol{T}_{\text{baseline}}(\boldsymbol{n}, \boldsymbol{N}) = \boldsymbol{c}_{\text{KEM}} + \boldsymbol{c}_{\text{AES}} \cdot \boldsymbol{N} + \boldsymbol{n} \cdot \boldsymbol{c}_{\text{Sig}} \tag{6}$$

The number of signature operations in $\Pi$ is exactly two: one Ed25519 signing call and one ML-DSA-65 signing call, regardless of $n$. This is the $O(1)$ signature-operation count that constitutes the central complexity contribution. The baseline performs $n$ dual signatures, giving an $O(n)$ signature-operation count.

### 4.4 Comparison

The cost saving of $\Pi$ over the baseline is:

$$\Delta T(n, N) = T_{\text{baseline}} - T_{\Pi} = (n-1) \cdot c_{\text{Sig}} - c_{\text{H}}(N + 32n) \tag{7}$$

$c_{\text{Sig}}$ which is a dual classical and lattice signature is substantially larger than the per-byte SHA3-256 hashing rate on the test platform. The savings term $(n-1) \cdot c_{\text{Sig}}$ dominates the additional hashing term $c_{\text{H}}(N + 32n)$ for all $n \geq 2$. Equivalently, $\Pi$ replaces $n$ per-file dual signatures with a single dual signature over a 32-byte commitment, plus one additional pass of SHA3-256 hashing over the batch ciphertexts.

Section 5.2 reports the measured signing-phase time reduction across batch sizes $n \in \{10,50,100,500,1000\}$ at three different file sizes (100 KB, 1 MB, and 10 MB). At $n = 1000$, the measured ratio of baseline to proposed signing-phase time ranges from 606-fold to 725-fold across the three file sizes, with the linear scaling of the baseline confirmed at $R^2 > 0.99$ in each case.

## 5. EXPERIMENTAL RESULTS AND PERFORMANCE ANALYSIS

### 5.1 Experimental Setup

All experiments were conducted on a single laptop platform which is an Intel Core i5-6300U CPU at 2.40 GHz (2 physical cores, 4 logical processors) with 16 GB of installed physical memory.

The benchmarks measure the signing-phase time from the start of the batch commitment construction or the first per-file signature call in the baseline to the completion of the final signature call. This is done across two parameters: batch size $n \in \{10,50,100,500,1000\}$ and file size $|m_i| \in \{100\text{ KB}, 1\text{ MB}, 10\text{ MB}\}$. Each combination is measured across 20 independent repeats to provide statistical stability. The signing phase is measured separately from bulk AES-256-GCM encryption to isolate the contribution of the commitment-based construction, which affects only the signing phase.

The proposed model $\Pi$ is compared against a per-file signing baseline that uses the same primitives, X25519, ML-KEM-768, HKDF-SHA-256, AES-256-GCM, Ed25519, and ML-DSA-65, but authenticates each file with its own dual signature rather than a single dual signature over a batch commitment. The only structural difference between baseline and proposed model is the SHA3-256 batch commitment $h_c$ and its associated single dual signature.

The implementation leverages Python 3 and uses three cryptographic libraries. The cryptography library provides X25519 elliptic-curve Diffie-Hellman key exchange, Ed25519 signing and verification, and HKDF-SHA-256 key derivation. The pycryptodome library provides AES-256-GCM authenticated encryption and SHA3-256 hashing for the batch commitment. The pqcrypto library provides ML-KEM-768 encapsulation and decapsulation and ML-DSA-65 signing and verification.

### 5.2 Results and Analysis

To verify the predicted reduction in signature-operation count, we measured the dual-signing time of the proposed scheme against a per-file signing baseline across batch sizes $n \in \{10,50,100,500,1000\}$ at a fixed file size of 1 MB. The per-file signing baseline performs n independent dual signatures, computing one Ed25519 signature and one ML-DSA-65 signature for each file. The proposed model performs a single dual signature over the SHA3-256 batch commitment $h_c$, regardless of $n$. In both schemes, the AES-256-GCM encryption and per-file ciphertext hashing are identical and are excluded from the timed region; only the signing phase is measured. Each configuration was timed twenty (20) times and the mean across repetitions were recorded.

***Table 2:*** *Signing Time for a File Size of 100KB*

| Batch Size (n) | Proposed Model (s) | Baseline Model (s) |
|---|---|---|
| 10 | 0.001198 | 0.010493 |
| 50 | 0.001192 | 0.056332 |
| 100 | 0.001925 | 0.107445 |
| 500 | 0.001471 | 0.488285 |
| 1000 | 0.001627 | 1.022693 |

***Table 3:*** *Signing Time for a File Size of 1MB*

| Batch Size (n) | Proposed Model (s) | Baseline Model (s) |
|---|---|---|
| 10 | 0.001040 | 0.010176 |
| 50 | 0.001107 | 0.050037 |
| 100 | 0.000952 | 0.102650 |
| 500 | 0.001363 | 0.512715 |
| 1000 | 0.001651 | 1.000424 |

***Table 4:*** *Signing Time for a File Size of 10MB*

| Batch Size (n) | Proposed Model (s) | Baseline Model (s) |
|---|---|---|
| 10 | 0.000917 | 0.009970 |
| 50 | 0.001157 | 0.049795 |
| 100 | 0.001184 | 0.100933 |
| 500 | 0.001364 | 0.493574 |

| Batch Size (n) | Proposed Model (s) | Baseline Model (s) |
|---|---|---|
| 1000 | 0.001352 | 0.980197 |

Tables 2-4 report the mean signing-phase time of the proposed and baseline models across batch sizes $n \in \{10, 50, 100, 500, 1000\}$, for per-file sizes of 100 KB, 1 MB, and 10 MB respectively, each averaged over 20 repetitions. In every case the baseline grows in direct proportion to $n$, rising from approximately 10 ms at $n = 10$ to close to 1.0 s at $n = 1000$, which is consistent with issuing one dual signature per file. The proposed model, which issues a single dual signature over the batch commitment, remains on the order of 1-2 ms throughout, varying only weakly and non-monotonically with $n$ rather than growing with it. At $n = 1000$ the resulting reduction in signing time is 629x for 100 KB, 606x for 1 MB, and 725x for 10 MB. These results confirm that the baseline cost is governed by the number of per-file signature operations, whereas the proposed model holds the number of post-quantum signature operations constant at one per batch irrespective of $n$, yielding a near-constant signing time whose residual variation reflects the transcript growing with batch size rather than any dependence of the signature count on $n$.

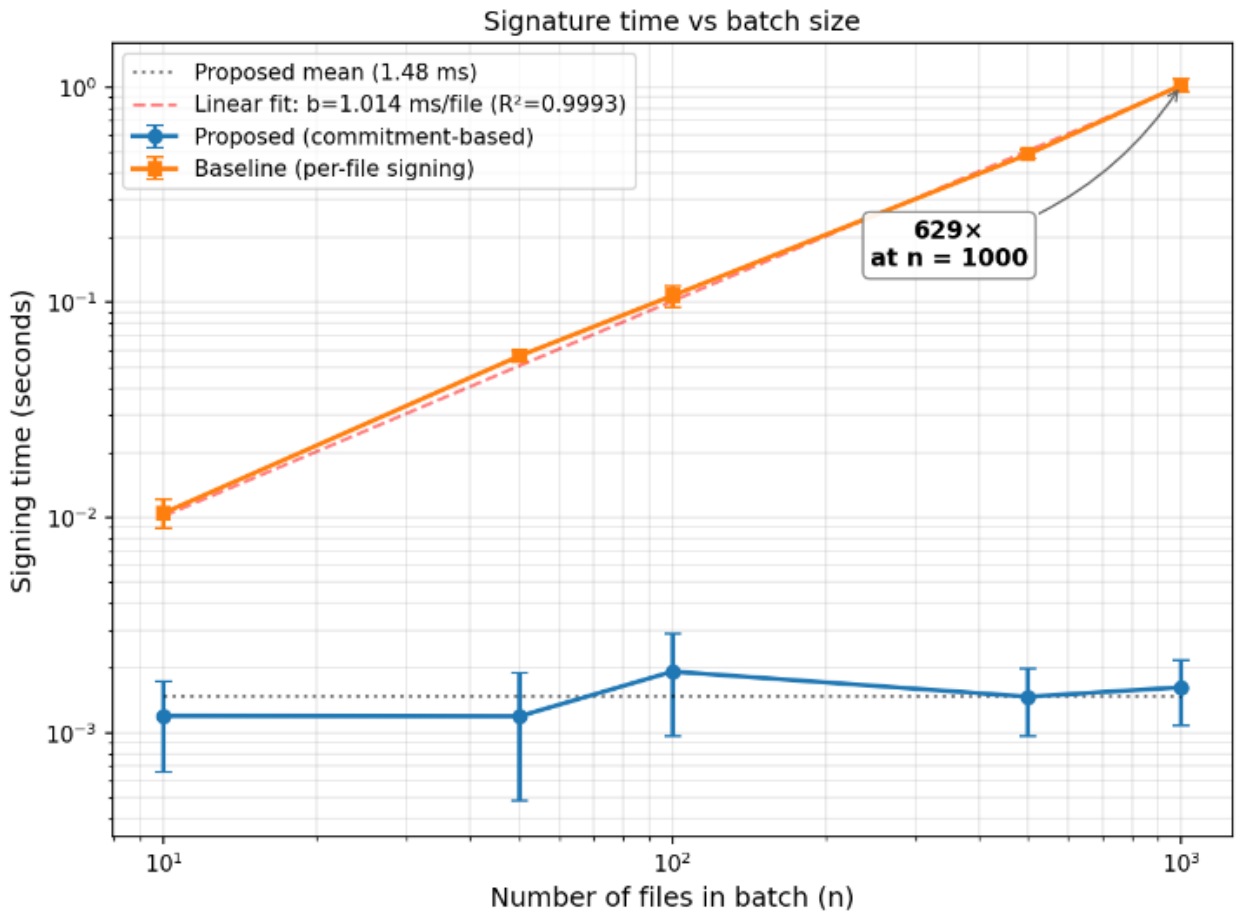

***Fig.4:*** *Signature Time vs Batch Size at 100KB per File*

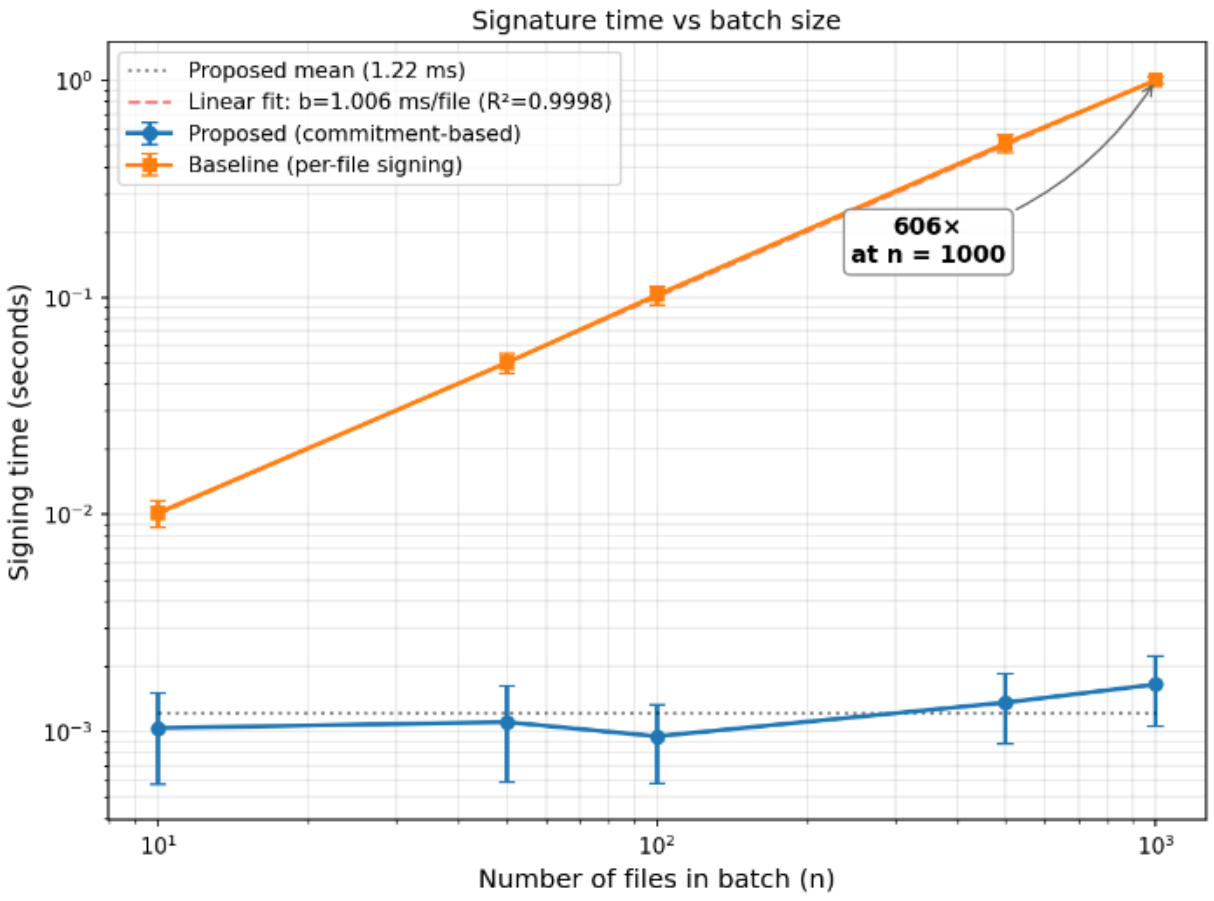

***Fig.5:*** *Signature Time vs Batch Size at 1MB per File*

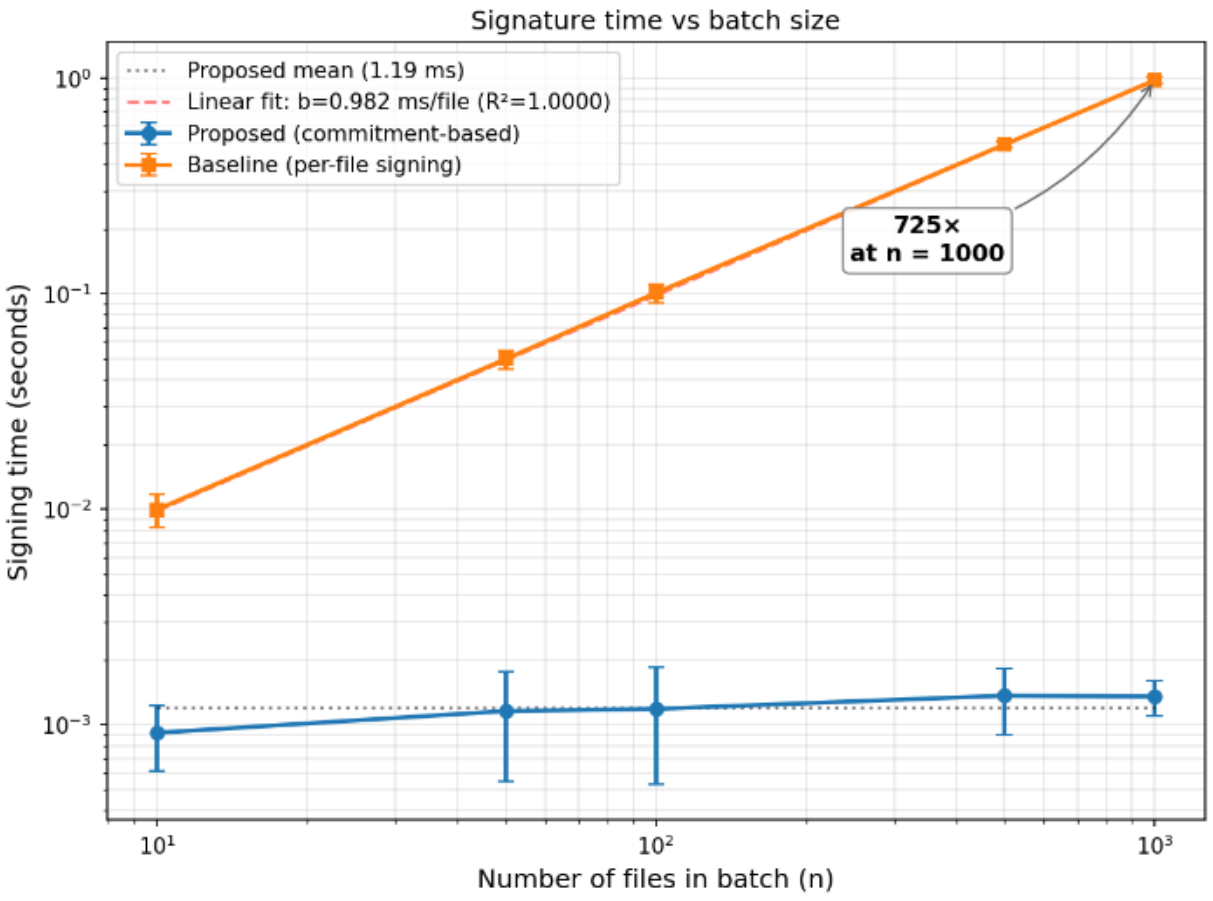

***Fig.6:*** *Signature Time vs Batch Size at 10MB per File*

Figures 4-6 show signing-phase time against batch size $n$ on log-log axes, for per-file sizes of 100 KB, 1 MB, and 10 MB. In all three, the baseline grows linearly with $n$, its through-origin fit yielding a per-file cost of approximately 1.0 ms/file with $R^2 \geq 0.999$, and rising from about 10 ms at $n = 10$ to roughly 1.0 s at $n = 1000$. The proposed model remains near-constant across the same range, with a mean of 1.19-1.48 ms depending on file size and only weak, non-monotonic variation in $n$ which is consistent with a fixed number of signature operations rather than a per-file cost. The reduction of signing time at $n = 1000$ is 629x for 100 KB, 606x for 1 MB, and 725x for 10 MB. The error bars on the proposed points are wide relative to its own sub-millisecond scale but negligible against the baseline, and the residual variation reflects the signed transcript growing with $n$ rather than any growth in the signature count, which remains fixed at two operations per batch.

## 6. LIMITATIONS AND FUTURE WORK

The empirical evaluation characterizes the sender-side signing-phase cost only. The corresponding receiver-side cost of reconstructing the SHA3-256 batch commitment, verifying both signatures, and decrypting each ciphertext is asymmetric with the sender-side cost and is not measured. Empirical characterization of this asymmetry across representative cloud retrieval workloads is a target for future measurement.

Secondly, the measurements were obtained on a single commodity client platform (Intel Core i5-6300U) using production Python cryptographic libraries. Evaluation on server-grade CPUs, constrained edge devices, and platforms with dedicated cryptographic acceleration would clarify how the reported improvements scale across deployment platforms and quantify the extent to which the wall-clock times are constrained by implementation-level rather than algorithmic factors.

Finally, the flat single-level commitment does not support selective per-file verification, which limits applicability in scenarios where a client retrieves a small subset of a larger batch. Alternative commitment structures, including Merkle-tree and Verkle-tree variants could be explored in the future.

## 7. CONCLUSION

This paper presented a commitment-based hybrid post-quantum cryptographic model for authenticated multi-file upload to cloud storage. The model comprises AES-256-GCM for bulk encryption, a hybrid X25519 with ML-KEM-768 for key encapsulation, and a hybrid Ed25519 with ML-DSA-65 dual signature, over a SHA3-256 batch commitment computed before signing. The distinguishing feature of the model is that this pre-signing commitment reduces the number of ML-DSA-65 signature invocations per batch from $n$ to one, independent of batch size, while the remaining work, which is bulk symmetric encryption and hashing, is bounded by fast symmetric primitive throughput.

Empirical evaluation on a commodity client platform confirms the structural claim. At a batch size of $n = 1000$, the proposed model reduces signing-phase time by a factor of 629, 606, and 725 for file sizes of 100 KB, 1 MB, and 10 MB respectively, relative to a per-file dual-signing baseline sharing every other primitive. Across all three file sizes the baseline grows in direct proportion to $n$ while the proposed model remains near-constant, its residual variation reflecting the signed transcript growing with batch size rather than any dependence of the post-quantum signature count, which stays fixed at two operations per batch, on $n$.